# Evolving Modular Genetic Regulatory Networks with a Recursive, Top-Down Approach

Javier Garcia-Bernardo* and Margaret J. Eppstein[†]

Department of Computer Science, University of Vermont, Burlington, VT, USA 05405

* Corresponding author: Javier.Garcia-Bernardo@uvm.edu
[†] Maggie.Eppstein@uvm.edu

Being able to design genetic regulatory networks (GRNs) to achieve a desired cellular function is one of the main goals of synthetic biology. However, determining minimal GRNs that produce desired time-series behaviors is non-trivial. In this paper, we propose a 'top-down' approach to evolving small GRNs and then use these to recursively boot-strap the identification of larger, more complex, modular GRNs. We start with relatively dense GRNs and then use differential evolution (DE) to evolve interaction coefficients. When the target dynamical behavior is found embedded in a dense GRN, we narrow the focus of the search and begin aggressively pruning out excess interactions at the end of each generation. We first show that the method can quickly rediscover known small GRNs for a toggle switch and an oscillatory circuit. Next we include these GRNs as non-evolvable subnetworks in the subsequent evolution of more complex, modular GRNs. Successful solutions found in canonical DE where we truncated small interactions to zero, with or without an interaction penalty term, invariably contained many excess interactions. In contrast, by incorporating aggressive pruning and the penalty term, the DE was able to find minimal or nearly minimal GRNs in all test problems.

## 1. INTRODUCTION

A genetic regulatory network (GRN) is a collection of genes that interact with each other and with the environment to govern the expression levels of proteins that help to control cell functions. Genes are encoded in the DNA and produce messenger RNA (mRNA) by a process called transcription, while proteins are produced from mRNA by a process called translation. The rate of transcription can be controlled by the binding of proteins, called transcription factors (TFs), to a gene's promoter region, which is a region of DNA needed to initialize the transcription of the gene. TFs can act as repressors or activators, decreasing or increasing the transcription rate of mRNA, respectively, when bound to the promoter region (Alon 2006). GRNs produce rich, complex behaviors that, when working together, regulate every biological organism. Genetic networks can be mathematically modeled, and the expression patterns recreated computationally (2004, de Jong 2002).

The field of synthetic biology aims to design and construct biological components with the goal of controlling cellular behavior to achieve a specific function (Mukherji and van Oudenaarden 2009). It is a rapidly growing field with many applications including biofuel production (Dunlop 2011), biological waste management (Gilbert, Walker, and Keasling 2003), and bio-sensing (Rajendran and Ellington 2008), among many others (Khalil and Collins 2010). Synthetic GRN circuits can be constructed and tuned to achieve a variety of desired functions (Khalil and Collins 2010). However, designing DNA circuits that can achieve a desired dynamical behavior can be a non-trivial task. Building a GRN requires identifying a large number of parameters that represent interactions between genes and proteins, transcription, translation and degradation rates. Evolutionary algorithms are stochastic global search algorithms that enable one to search this large space of parameters for combinations that produce a desired output. In an evolutionary algorithm, the fitness of a set of candidate solutions (individuals, where each solution contains different parameters of the model) is calculated, and the best performing individuals are combined in subsequent generations in order to find optimal or near-optimal solutions. Since François started evolving GRNs in 2004 (François and Hakim 2004), researchers have been using evolutionary algorithms to evolve network motifs or synthetic networks with a desired behavior (François and Hakim 2004, Noman, Palafox, and Iba 2013b, Drennan and Beer 2006, van Dorp, Lannoo, and Carlon 2013b, van Dorp, Lannoo, and Carlon 2013a); see (Noman, Palafox, and Iba 2013a) for a recent review. However, those studies rarely evolved minimal or near-minimal GRNs. Even in methods that attempted to prune away excess parts of the network (Drennan and Beer 2006, van Dorp, Lannoo, and Carlon 2013b), the number of interactions of the evolving networks increased with the number of generations.

Synthetic biology is also used to help scientists understand how natural biological systems are genetically assembled and how they operate *in vivo* (Mukherji and van Oudenaarden 2009). The ability to understand the behavior of small GRNs makes it easier to design and produce synthetic circuits that will behave in a predictable manner. Some small biological circuits, such as bistable and oscillatory circuits, are well-understood. Bistable GRNs, those that can exhibit two mutually exclusive states depending on initial conditions, are ubiquitous in nature. For instance, stem cells can remain undifferentiated for years and differentiate when needed, moving from one stable state to another (Wang et al. 2009). A minimal GRN for a bistable switch with monomeric factors is well known (Widder, Macía, and Solé 2009). Similarly, oscillatory GRNs, those whose genes are expressed in cycles, are found in every plant and animal and many bacteria, controlling global behaviors including sleep, stress and light responses (Levine, Lin, and Elowitz 2013). For example, the protein p53 oscillates when mammal cells are exposed to UV or gamma radiation (Purvis and Lahav 2013). The so-called "repressilator" is a well-known minimal GRN that produces oscillations (Elowitz and Leibler 2000). However, the networks for many natural GRNs are not known or fully understood. The development of single-cell microscopy and high-throughput gene expression analysis has allowed the study of simultaneous expression levels of hundreds of genes under different conditions, and GRN inference from gene expression levels is active area of research (Marbach et al. 2012, Marbach et al. 2010, Bar-Joseph, Gitter, and Simon 2012) where evolutionary computation is proving useful (Cao et al. 2010, Ruskin and Crane 2010). While our primary motivation is for designing GRNs for synthetic biology, the algorithm introduced in this work could be used for the inference of naturally-occurring GRNs from gene expression data.

In this contribution we use differential evolution (DE), an evolutionary algorithm that is explicity designed for efficient evolution of candidate solutions that are represented as vectors of real-valued variables (Storn and Price 1997). Using DE, we evolve small GRNs with four different types of desired time-series behaviors. We experiment with three different approaches for minimizing excess interactions between genes during the evolution. We first assess how well these methods can evolve known minimal GRNs that produce bistable or oscillatory behaviors. We then show how one can incorporate previously identified GRNs as non-evolvable subnetworks into the subsequent evolution of larger (but still minimal or nearly-minimal) modular GRNs for two more complex types of dynamical behaviors. We find that a top-down approach, where we start with many more interactions than needed, enables us to rapidly identify a non-minimal network that contains a desired dynamical behavior. Once the target behavior is detected, we switch the strategy of the DE to a more focused search while also pruning away up to five interactions per candidate solution per generation and increasingly penalizing for the number and strength of interactions. Using this approach we are able to evolve small, often minimal, GRNs. In contrast, the methods that did not incorporate aggressive pruning resulted in GRNs that invariably contained many excess interactions, whether or not fitness was increasingly penalized for the number and strength of interactions.

## 2. MATERIALS AND METHODS

### 2.1. Model of gene expression

GRNs are commonly modeled as systems of differential equations with degradation and production terms (Ruskin and Crane 2010, Smolen, Baxter, and Byrne 2000). Promoter dynamics and mRNA concentrations are usually modeled with Hill functions, as shown in Eq. (1), while protein concentration depends on the concentration of mRNA as shown in Eq. (2). Normalizing mRNA and protein concentrations, the pair of differential equations corresponding to expression for each gene $i$ can be expressed as

$$\frac{dm_i}{dt} = -m_i + \frac{\alpha_i}{1 + K_{1,i}p_1^{n_{1,i}} + K_{2,i}p_2^{n_{2,i}} + \ldots + K_{G,i}p_G^{n_{G,i}}} + \alpha_0 \quad (1)$$

$$\frac{dp_i}{dt} = -\beta_i(p_i - m_i) \quad (2)$$

where $m_i$ is mRNA level for gene $i$, $p_i$ is protein level of the protein transcribed from gene $i$, $\alpha_i$ is the maximum expression of the promoter, $\beta_i$ is the translation rate, $\alpha_0$ is the basal level of gene expression, and $G$ is the number of potentially interacting genes (Elowitz and Leibler 2000). Together, $K_{i,j}$ and $n_{j,i}$, determine the strength of an interaction. Each $n_{j,i}$ determines the sign and strength of the interaction between the protein $j$ and the promoter region of the gene $i$ that turns on expression of $m_i$. and $K_{j,i}$ is the strength of repression/activation once the protein is bound to the promotor. When $n_{j,i}$ is zero, there is no feedback of protein $p_j$ on gene $i$ and therefore no interaction. Positive $n_{j,i}$ indicate that the protein $p_i$ is a repressor or the gene $i$. This representation of gene expression accurately models the biological system and has been broadly used (Smolen, Baxter, and Byrne 2000), but it requires many parameters.

Using this model of gene expression for simulating the dynamical behavior of a GRN with $G$ genes one must thus solve $2 \cdot G$ coupled differential equations. In this work, we used Matlab's *ode15s* to solve these potentially stiff systems and obtain a time series of simulated protein expression.

For a GRN with $G$ genes there are potentially $G^2$ interactions, each with two associated coefficients ($n_{j,I}$ and $K_{j,i}$), and each gene has two other associated coefficients ($\alpha_i$ and $\beta_i$). Thus, for the inverse problem of evolving GRNs, there are $M \leq 2 \cdot G^2 + 2 \cdot G$ unknown decision variables that must be identified to fully characterize the GRN (the inequality is because some of the potential interactions may be fixed and not subject to evolution).

**2.2. Evolutionary algorithm**
In preliminary experimentation with two evolutionary algorithms: covariance matrix adaptation evolution strategy (CMA-ES (Hansen 2006)) and differential evolution (DE (Storn and Price 1997)), DE proved to be able to handle boundaries more easily and in general performed better for this application. Thus, all reported results here used DE (we used the open-source Matlab implementation of DE available here: http://www1.icsi.berkeley.edu/~storn/code.html#matl). DE evolves a population of $N$ solutions, each of which is represented as a real-valued individual vector of length $M$.

Crossover rate was set at the default value of 0.8. We then implemented the following modifications. We performed up to three restarts when the fitness had not improved for ten generations (note that we did count fitness evaluations of aborted starts in the total for the trial run). Evolutionary runs started with the DE/rand/1/bin strategy, where each mutant solution ($\vec{v}_i$) was created by adding the difference of two random population members ($\vec{x}_{r_2}$ and $\vec{x}_{r_3}$) scaled by a factor $F$, to another *random* population member ($\vec{x}_{r_1}$), as shown below. $\vec{x}_i$ was replaced by $\vec{v}_i$ when the former vector had an increased fitness:

$$\vec{v}_i = \vec{x}_{r_1} + F(\vec{x}_{r_2} - \vec{x}_{r_3}) \qquad (3)$$

As soon as the desired dynamical behavior was found in a solution (when raw fitness fell below a pre-specified threshold; see Section 2.4 *"Raw fitness functions"* for details), the strategy was changed to DE/rand-to-best/1/bin, where each mutant solution was created by adding the scaled difference between this *best* population member and a random population member to another scaled difference vector of two other random population members, as shown below:

$$\vec{v}_i = F(\vec{x}_{best} - \vec{x}_{r_1}) + F(\vec{x}_{r_2} - \vec{x}_{r_3}) \qquad (4)$$

At the same time as this strategy shift occurred, we also changed the DE scaling parameter $F$ from 0.8 (default parameter) to 1.2. The strategy shift enabled the DE to focus the search around the best GRN to speed up the evolution, but increasing $F$ also allowed larger step sizes to permit exploration of other nearby topologies that may be better.

**Table 1. Coefficient ranges.**

| Method | Penalty Term | Truncation of small values | Aggressive Pruning |
|---|---|---|---|
| *ForcedReduction* | **yes** | no | **yes** |
| *NoPenalty* | no | **yes** | No |
| *WithPenalty* | **yes** | **yes** | No |

Each of the decision variables was represented as a bounded real variable, uniformly randomly initialized from the ranges shown in Table 1. Following random initialization, some of the $G^2$ Hill coefficients $n_{j,i}$, were then randomly set to zero to control the density of the initial interconnection network. In this work, we tried evolving GRNs starting from three different levels of initial interconnection density: *dense*, *medium*, and *sparse* (see Section 2.3 *"Test problems"* for details.)

As the evolution progressed, parameter boundaries were checked and values outside the allowable range were truncated to the maximum (or minimum) of the ranges specified in Table 1. For all experiments reported here, $\alpha_0 = 0.2$. These ranges for the system parameters are in agreement with previous studies (Elowitz and Leibler 2000, Kim, Yoon, and Cho 2008), but could easily be modified to evolve other systems with values outside them (e.g., if one desired an ultrasensitive switch, where the Hill coefficient $n_{j,i}$ is larger than 3).

Since the goal of this study was to evolve minimal GRNs, we tried various ways of reducing excess interactions in the evolving solutions. These included (i) penalizing for the number and strength of interactions, (ii) aggressively pruning out interactions, and (iii) truncating small interaction coefficients to zero. In this paper, we report on three methods that used various combinations of these techniques, as follows.

In two of the methods reported on here, we included the following Penalty term:

$$Penalty = \frac{gen}{C} \sum_i \sum_j |n_{j,i}| \qquad (5)$$

The penalty is proportional to the sum of the absolute values of Hill coefficients $n_{j,i}$, thus penalizing for the total number and strength of interactions. The penalty term is multiplied by the generation number *gen*, divided by a positive constant $C$, so that the penalty would increase slowly as the evolution progressed. This constant was set to allow for small contributions to the fitness at early generations and penalties contributing up to 20% at later generations. As a rule of thumb, the expected final number of interactions times the maximum number of generations should equal 10% of the optimal fitness. Because we were minimizing fitness, the *Penalty* term was added to the raw fitness (see Section 2.4 *"Raw fitness functions"* for details).

*ForcedReduction Method:* In the main method we are proposing, we used the *Penalty* term along with an aggressive pruning step at the end of each DE generation subsequent to the strategy shift to rand-to-best, as follows. For each solution that exhibited the desired dynamics fitness, we iterated through each Hill coefficient $n_{j,i}$, temporarily set the coefficient to zero, and if the raw fitness of the modified solution was no more than 15% worse than the raw fitness of the solution with that coefficient, the zero was kept in the solution vector. i.e., we accepted as much as 15% degradation in fitness to increase parsimony. This process was continued until we had replaced five coefficients with zeros, we ran out of Hill coefficients to try or the fitness of the modified solution was 10% worse than the unmodified one (we accepted a 15% degradation in an individual step but only a 10% global degradation). Preliminary studies showed that these two thresholds worked well in combination. Note that higher thresholds increase the aggressiveness of the method, and therefore if thresholds are too high it may contribute to premature convergence to a suboptimum. In contract, lower thresholds decrease the aggressiveness and therefore decrease the speed of the algorithm and the ability to reduce the number of interactions.

**Table 2. Techniques to reduce interactions used in this paper.**

| Coefficient | Initial range | Bounded range |
|---|---|---|
| $n_{j,i}$ | [-3,3] | [-3,3] |
| $K_{i,j}$ | [1,2] | [0.5,5] |
| $\alpha_i$ | [100,500] | [0.5,500] |
| $\beta_i$ | [0.5,5] | [0.5,5] |

*NoPenalty Method:* For comparison, we implemented a second method in which we ran the DE without using the *Penalty* term and without any aggressive pruning. In this method, we simply truncated very small Hill coefficients $n_{j,i}$ (those with absolute values less than 0.5) to zero. This value was chosen to disrupt weak interactions and create parsimonious, stable networks.

*Penalty Method*: Finally, we implemented a third method that was exactly like the *NoPenalty* method except that it also used both the *Penalty* term from Eq. (5).

We summarize the combinations of techniques incorporated into these three methods in Table 2.

Storn and Price (Storn and Price 1997) recommend DE population sizes of between 5-10 individuals per decision variable. However, since we had so many decision variables (for most problems tested here we had 84 unknowns), and since our fitness function requires numerical integration and so is relatively slow, we opted to use much smaller population sizes of only $N = 25$ individuals, which was found to be sufficient during preliminary experimentation. The only exception to this was for our most difficult test problem (*DualOscillator*, described in Section 2.3 *"Test Problems"*), in which we started with a population of $N = 50$ individuals, but then discarded all but the best 25 individuals as soon as the desired dynamical behavior was found and continued the evolution with these $N = 25$ remaining individuals. We recommend 5-10 individual per expected final number of interactions. Values of problem-specific parameters are summarized in Table 3.

**Table 3. Parameters of the four test problems.**

| Test Problem | DE popsize $N$ | Penalty Divisor $C$ | # GRN Genes G | # decision variables $M$ |
|---|---|---|---|---|
| *Bistable* | 25 | 250 | 6 | 84 |
| *Oscill.* | 25 | 250 | 6 | 84 |
| *Cond. Oscill.* | 25 | 250 | 3 evolved + 5 fixed | 36 |
| *Dual Oscill.* | 50 to 25 | 2500 | 6 evolved + 3 fixed | 84 |

All DE trials were started from one of three levels of initial network interconnection density (*dense*, *medium*, and *sparse*); the initial number of non-zero interactions prescribed for each of these three densities, for each of the four test problems, are shown in Table 4.

### 2.3. Test Problems

In this study, we first tested each approach to see how well we could evolve known small GRNs for a simple toggle switch (*Bistable*) and an oscillatory circuit (*Oscillator*). We then included one *Bistable* GRN and/or one *Oscillator* GRN as non-evolvable subnetworks in the subsequent evolution of more complex GRNs, to illustrate how the methods can be recursively applied to bootstrap the evolution of increasingly complex, modular GRNs. In one problem, we evolved GRNs that combined non-evolvable subnetworks that included a toggle switch and an oscillatory circuit (*ConditionalOscillator*), as in (Thomas and Jin 2012); to make this problem more difficult we precluded direct interactions between these two subnetworks. In the other problem, we evolved GRNs with two mutually exclusive oscillators (DualOscillation); to make this problem more difficult we only included one copy of a non-evolvable oscillatory circuit.

**Table 4. Number of non-zero Hill coefficients $n_{j,i}$ (interactions) in initial solution vectors for the three initialization densities.**

| Test Problem | *Dense Initial Network* | *Medium Initial Network* | *Sparse Initial Network* |
|---|---|---|---|
| *Bistable* | 33 | 18 | 4 |
| *Oscill.* | 33 | 18 | 4 |
| *Cond. Oscill.* | 13 | 8 | 3 |
| *Dual Oscill.* | 33 | 18 | 4 |

### 2.4. Raw Fitness Functions

Assessing whether or not a GRN exhibits a desired dynamical behavior requires using the GRN to simulate time-series data and then computing some metric (referred to as the raw fitness) for how closely that time-series data meets the desired characteristics. Appropriate raw fitness metrics are dependent on the nature of the particular target dynamical behavior. Thus, each of the four test problems had their own unique raw fitness metric, as described below.

*Bistable raw fitness:* To detect bistability, raw fitness was calculated as (the negative of) the absolute difference in the mean levels of one pre-specified target protein, between two 50-time step simulations. In one simulation the initial levels of the target protein were set to 1, in the other simulation the initial levels were set to 100. The minimum value of the raw fitness was truncated at -15, so as not to reward GRNs that included excess positive regulation of one of the proteins by extra genes. Such excess positive regulation is not necessary for bistability, but does increase the difference between the two protein levels. It is worth noting that the difference between two stables states can thus be easily increased to a desired level by adding in extra activators. When the raw fitness of an evolving solution fell below the threshold of -9, the correct behavior for the *Bistable* circuit was considered to be found, triggering the strategy shift described in Section 2.2 *"Evolutionary algorithm"*.

The other three test problems (*Oscillator*, *ConditionalOscillator*, and *DualOscillator*) each contained an oscillatory component. The presence of oscillatory behavior was assessed using an autocorrelation-based fitness metric, as follows. Auto-correlation is the correlation between a time series and the same time series with a specified time-lag. Here, we simulated time series data from evolving GRNs, starting from initially random protein levels, for 100 time steps; we then computed a vector of normalized unbiased auto-correlations of target oscillatory protein, for time lags from 0 to 50 time steps, using Matlab's *xcorr* function. A perfectly oscillatory GRN will have a periodic unbiased auto-correlation vector that oscillates between -1 and +1. The autocorrelation-based fitness metric was calculated as the sum of the first local minimum in this autocorrelation vector plus two times the sum of the second through fifth local minima (in the positive domain), where the minima are found using the Matlab *findpeaks* function. The first minimum is weighted less to account for an initial transient period before the protein levels settle into oscillatory behavior, but the first minimum is not completely ignored to account for the case where there is only one minimum. Thus, for a perfectly periodic time series with no transient period, this autocorrelation-based fitness metric will be -9; damped oscillatory behaviors will have a smaller (but still negative) magnitude. Details of how the specific raw fitnesses for the oscillatory circuits were calculated from this autocorrelation-based fitness metric follow.

*Oscillatory raw fitness:* Because the *Oscillator* behavior was found easily with our algorithm, the raw fitness was taken to be the autocorrelation-based fitness measure for one arbitrarily chosen protein. When the raw fitness fell below the threshold of -5.5, the correct behavior for the *Oscillator* circuit was considered to be found, triggering the strategy shift described in Section 2.2 *"Evolutionary algorithm"*. This threshold depends on the problem considered and sets the timing of the algorithm, with higher (less negative) thresholds allowing for longer searching times before the strategy shift.

*ConditionalOscillator raw fitness:* To assess the raw fitness of the *ConditionalOscillator* GRN solutions, two different simulations were run: in the first simulation, the protein and mRNA levels of one of the proteins in the *Bistable* building block were initially set to 1; in the second simulation they were both set to 100. The raw fitness was computed as the difference in the autocorrelation-based fitness measures of the first and the second simulation. When the raw fitness fell below the threshold of -5.5, the correct behavior for the *ConditionalOscillator* circuit was considered to be found, triggering the strategy shift described in Section 2.2 *"Evolutionary algorithm"*.

*DualOscillator raw fitness:* To assess the raw fitness of the *DualOscillator* GRN solutions, two different simulations were run; in the first simulation, the hard-coded oscillatory building block were initially oscillating, and in the second simulations its protein levels were initially flat. The raw fitness was computed as the difference between the autocorrelation-based fitness measure of the first and second simulation. When the raw fitness fell below the threshold of -3, the correct behavior for the *DualOscillator* circuit was considered to be found, triggering the strategy shift described in Section 2.2 *"Evolutionary algorithm"*.

### 2.5. Experimental procedures

We conducted 25 repetitions of each the 3 methods on each of the 4 test problems, starting from each of 3 the different densities of interactions in the initial solution vectors, for a total of 900 experimental runs.

Evolutionary runs using *ForcedReduction* were terminated after a maximum of 50 generations; those using the *NoPenalty* or *Penalty* methods were terminated after a maximum of 100 generations to compensate for the fact that *ForcedReduction* had more fitness evaluations per generation when it was actively pruning excess interactions. However, runs were terminated earlier if the number of interactions in the best evolved GRN did not change for 5 generations when using *ForcedReduction* or for 30 generations when using the other two methods (because the reduction in the number of genes is slower in these), after the desired behavior was found. Our results show that *ForcedReduction* actually required fewer total evaluations to terminate than did either of the other methods, as shown in *"Results"*.

Autoregulation (where a gene has a positive or negative feedback on itself) is not necessary for any of the behaviors we sought, so because we were seeking minimal networks autoregulation was never spontaneously evolved. However, while not strictly necessary to achieve the desired behaviors, autoregulation can help to make certain GRNs more robust (Komiya, Noman, and Iba 2012). Thus, at the end of each evolutionary run, we tested each autoregulation Hill coefficient $n_{i,i}$ with the minimum value of -3 (positive autoregulation) and the maximum value of 3 (negative autoregulation); if either of these values improved the raw fitness then the altered value was kept in the final solution. This only proved beneficial in the case of *Bistable,* given our fitness criteria.

### 2.6. Success criterion

Following the termination of each experimental run, we determined whether or not the run was successful, where only GRNs that were robust in being able to generate the desired dynamical behavior were considered successful. To assess robustness, we performed 100 simulations (50 time steps for the *Bistable* GRNs and 100 time steps for the other three test problems) of the best resulting GRN from a given run. Each simulation started from random concentrations of mRNA and protein (selected from a uniform distribution between 0 to 20 molecules per species). We then assessed whether or not the target behavior was present in each of these simulations. If the number of successful simulations was at or above 90 out of 100, we considered the GRN robust and the run successful.

### 3. RESULTS

Of the 900 total runs, 689 (77%) met our criterion for success, with 422 exhibiting the desired dynamical behavior in all 100 of 100 simulations (Figure 1). Using the *ForcedReduction* method we were able to successfully evolve small, often minimal, GRNs that produced the four desired behaviors (Figure 2). For example, in the *Bistable* problem, successful solutions consistently recovered the well-known monomeric bistable switch (Widder, Macía, and Solé 2009), with only 2 evolved nodes and 2 evolved interactions. The positive autoregulatory interaction was added in the post-processing step to increase the robustness of the bistability (Figure 2a). Similarly, in the *Oscillatory* problem, successful solutions invariably found the well-known repressilator (Elowitz and Leibler 2000) shown in Figure 2b, with 3 evolved nodes and 3 evolved interactions. In contrast, in (Drennan and Beer 2006) the repressilator was found as the core of oscillatory genetic networks that were evolved, but those networks contained many excess interactions. We were also able to evolve small GRNs for the two more complex problems. In the *ConditionalOscillator* problem, most successful runs found the nearly-minimal GRN with 2 evolved nodes and 3 evolved interactions connecting the hardcoded bistable and oscillatory motifs, shown in Figure 2c. Results on the *DualOscillator* problem were slightly more variable; two successful evolved solutions that were commonly found are shown in Figure 1d, one with 5 evolved nodes and 7 evolved interactions, and one with only 3 evolved nodes and 5 evolved interactions.

Recall that, in the *Bistable, Oscillatory*, and *DualOscillatory* problems, we were starting from 6 evolvable nodes (so 36 evolvable interactions), with either 33, 18, or 4 of the interaction coefficients initially non-zero, whereas in the *ConditionalOscillator* problem, we were starting from 3 evolvable nodes (so 9 evolvable interactions), with either 13, 8, or 3 of the interaction coefficients initially non-zero (Table 4). The number of genes and interactions in successfully evolved GRNs using the *ForcedReduction* method was nearly insensitive to these initial number of non-zero interaction coefficients, and minimal or near-minimal GRNs were found in all cases (Figure 3a). The success rate, however, was lower when starting from the sparsest initial interconnection networks initially set to zero, as shown by the shorter white bars in Figure 3b. In all except the *Bistable* problem, we observed that, when starting from the sparsest initial networks, we often observed an evolved there was in initial increase in the number of interactions before the desired behavior was found, and then the size of the best GRN fluctuated as nearby topologies were explored and found to be better (solid red lines, Figure 4). Ultimately, the increasing penalty term helps *ForcedReduction* to prune away the excess interactions. In the difficult *DualOscillator* problem, this also occurred when starting from the initial networks with medium density (e.g., red solid lines, Figure 4b,c). Thus, runs starting from sparse networks took longer to find the desired behavior than did runs starting from the denser networks; this resulted in fewer successes and, in the case of the two more complex networks, sometimes slightly larger final GRNs (Figure 3a), although not significantly so ($p > 0.1$).

The *NoPenalty* method was also very successful in identifying GRNs with the desired dynamical behavior, with slightly higher success rates than *ForcedReduction* for 3 of the 4 test problems, and less sensitivity to the initial sparsity of the networks (compare Figure 3d to 3b). However, even though this method truncated small non-zero interaction coefficients to zero, all of the successfully evolved GRNs had many excess interactions, regardless of the initial sparsity of the network (compare Figure 3c to 3a, noting the difference in the scale of the y-axis). Note how, on the difficult *DualOscillator* problem, the number of interactions actually increased to about 35 (Figure 3c; blue dashed lines, Figure 4) regardless of the initial density of the network. The robust success rates of this approach indicate that perhaps a hybrid approach that starts with *NoPenalty* and then transitions to *ForcedReduction* may be better than either method alone, yielding high success rates and minimal GRNs, although we have not yet tried this.

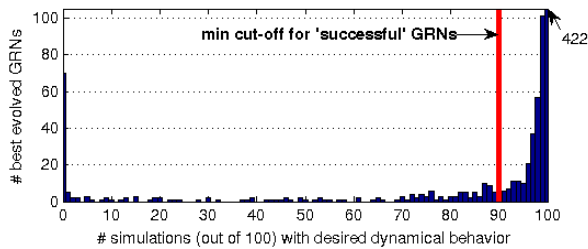

**Figure 1. Histogram of the number of successful simulations (out of 100) of the best evolved GRNs resulting from the 900 experimental runs.** Those with at least 90 successful simulations (above the red line) were considered 'successful' runs.

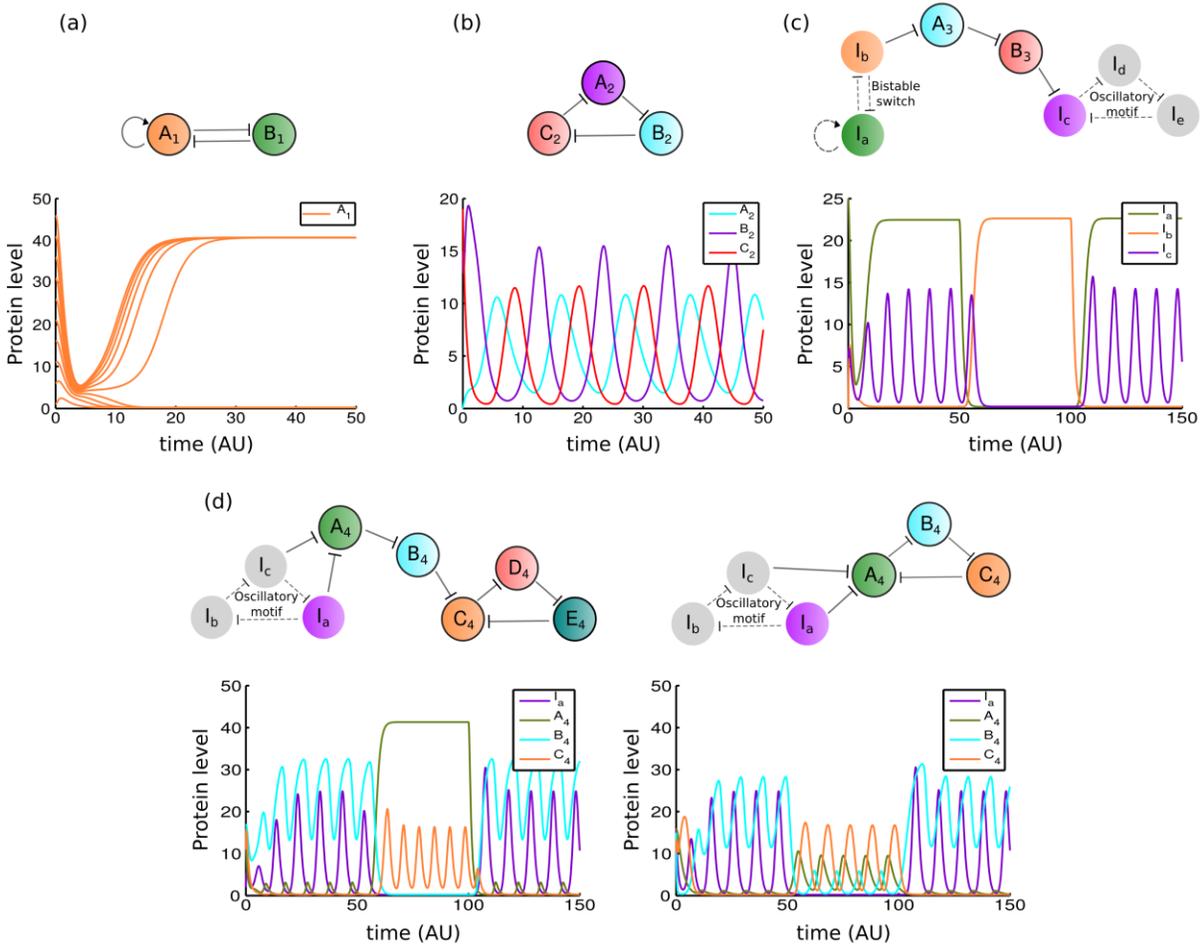

**Figure 2. Representative evolved small GRNs and their time-series behaviors.** Nodes represent proteins coded for by specific genes. Nodes outlined in black and interactions shown with solid lines were evolved. Nodes with no outline and interactions shown with dotted lines were included as non-evolvable subnetworks. All nodes and interactions were evolved in (a) *Bistable*, and (b) *Oscillator*, with the exception that the (non-essential) positive autocorrelation shown in *Bistable* was added in the post-processing phase. In (c) *ConditionalOscillator*, 2 nodes and 3 interactions were evolved such that protein $I_c$ oscillates when $I_a$ is present but not when $I_b$ is present (here, we externally manipulated $I_b$ at time steps 50 and 100). In (d) *DualOscillator*, two successful solutions are shown, one in which 5 nodes and 7 interactions were evolved, and one in which 3 nodes and 5 interactions were evolved. In both solutions, oscillations in protein $I_a$ and $C_4$ are mutually exclusive (here, we externally manipulated $I_c$ at time steps 50 and 100).

When the *Penalty* method was successful, the number of interactions in the evolved GRNs were lower than with the *NoPenalty* method (compare Figure 3e to 3c), showing that the penalty term was, indeed, helping to reduce the number of interactions. However, there were still many excess interactions in comparison to the small networks evolved with *ForcedReduction* (compare Figure 3e to 3a). Unlike in the other two methods, the number of interactions in successful GRNs evolved using the *Penalty* method was significantly smaller when starting from the sparest network in comparison to when starting from the densest network ($p < 0.001$ for all four test problems) because the penalty term slowed down the addition of extra interactions. However, this also made it harder for the *Penalty* method to find successful solutions, resulting in lower success rates than the *NoPenalty* method (compare Figure 3f to 3d). Unsuccessful runs occurred when the growing penalty term dominated the raw fitness value and removing interactions was favored over finding and retaining the desired dynamical behavior. To be fair, the dynamically growing penalty term was crafted to work in conjunction with the aggressive pruning of *ForcedReduction*, where there was no need to explicitly prevent the penalty term from dominating raw fitness. One could certainly incorporate such a safeguard for use in the *Penalty* method. However, even in the successful runs with *Penalty*, the method was not capable of removing many of the excess interactions. For example, on the *DualOscillator* problem the number of interactions in successful solutions was high (Figure 3e) and appears to have plateaued, with no sign of further decrease as the evolution progressed (black dash-dot lines, Figure 4). This indicates that a less dominant penalty term applied without aggressive pruning may improve the success rate of the *Penalty* method, but would still not be able to achieve the small GRNs evolved by *ForcedReduction*, even when allowed to run for many more generations.

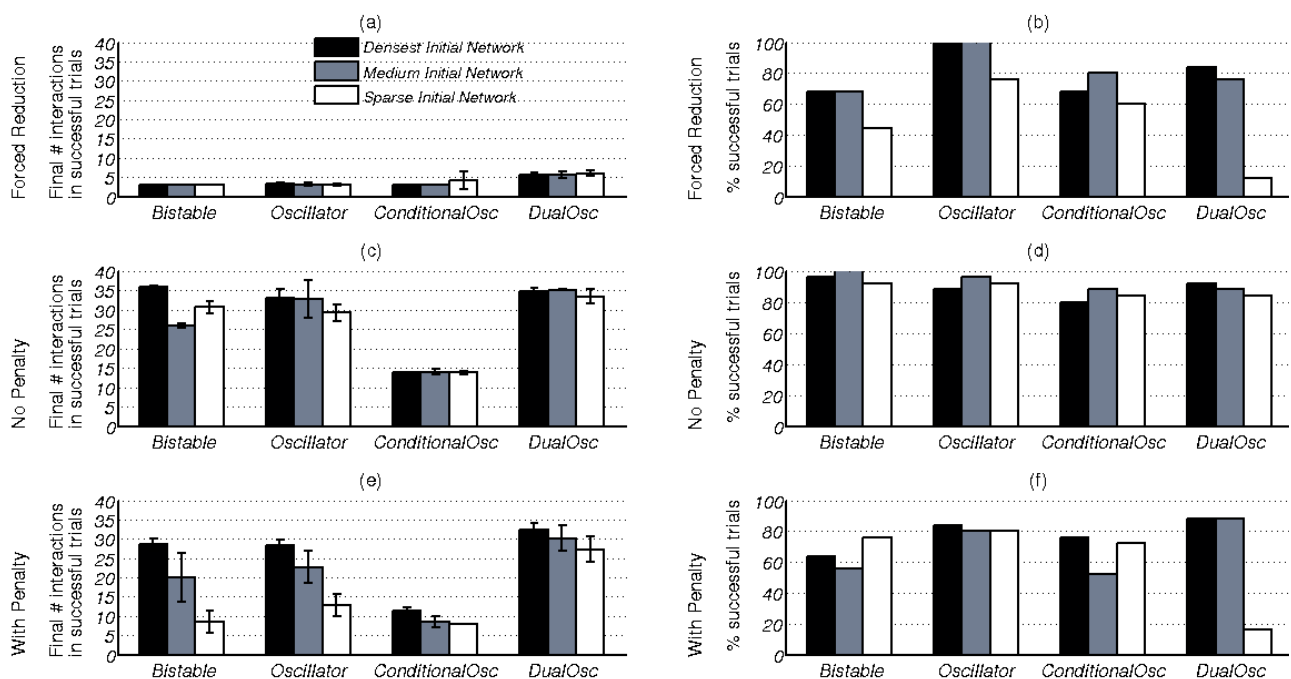

**Figure 3. Success rates and final number of interactions in successful trials for each method, initial network density, and test problem..** In the left column (a,c,e), the heights of the solid bars show the mean number of interactions in the final GRNs of successful runs and the error bars indicate one standard deviation; note the different scales on the y-axis. In the right column (b,d,f) we show the percentage of successful trials (out of 25). The first row (a,b) shows results for the *ForcedReduction* method, the second row (c,d) shows results for the *NoPenalty* method, and the third row (e,f) shows results for the *Penalty* method. The legend in (a) applies to all 6 plots.

## 4. DISCUSSION

Here, we have shown that starting from initially excessively dense interaction networks makes it relatively easy for DE to discover networks that exhibit the desired dynamical behaviors, even when using population sizes that are much smaller than are generally recommended. Excess interactions can then be aggressively pruned away as the evolution continues, even when this entails a temporary degradation in the dynamical behavior. This approach, in conjunction with a dynamic penalty term that increasingly penalizes for the number and strength of interactions as the GRNs evolve, often yields minimal, or nearly-minimal, GRNs that robustly exhibit the desired dynamical behavior when starting from a wide variety of initial conditions. Using this approach, we were able to consistently find well-known GRNs for the oscillatory repressilator (Elowitz and Leibler 2000) and a small bistable switch (Widder, Macía, and Solé 2009). Others who have tried to incorporate pruning of excess interactions (Drennan and Beer 2006, van Dorp, Lannoo, and Carlon 2013b) have not succeeded in evolving minimal networks.

We also demonstrated how one can recursively apply the method to evolve larger, more complex, modular GRNs. This is accomplished by incorporating previously discovered GRN circuits as non-evolvable subnetworks in subsequent runs, so that that the evolutionary process may utilize these circuit without having to rediscover them. Using this approach, we were able to evolve larger (but still small) GRNs that produced two types of more complex dynamical behaviors by combining oscillatory and/or bistable GRNs. Because we are evolving gene interaction coefficients, the number of decision variables grows quadratically with the number of evolvable genes. However, by recursively using the method to incrementally build up more complex networks, the DE can focus on only a relatively small number of evolvable genes and interconnections during a given run, thus keeping the method tractable.

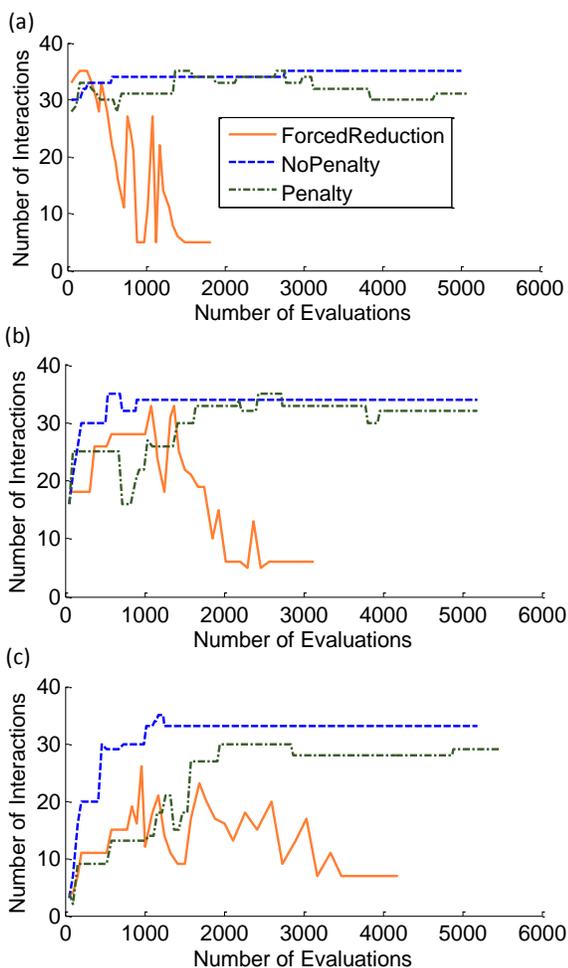

**Figure 4. Number of interactions in the best current population member during the evolution for three representative successful runs on the *DualOscillator* problem and three initial network densities.** (a) the densest initial networks, (b) initial networks of medium density, and (c) the sparsest initial networks. The legend in the (a) applies to all three plots. Note that the x-axis is number of evaluations, not generations, in order to fairly compare the three methods. In all cases, the final networks exhibited the desired dynamical behavior in 100 out of 100 random simulations.

We compared this approach to two more canonical versions of DE, using more traditional approaches for removing excess interactions. In both of these methods we truncated small interaction coefficients to zero; one of the methods used the same dynamic penalty term described above and the other used no penalty function. While these methods were often successful in finding GRNs with the desired dynamical behaviors, they invariably yielded non-minimal GRNS that included many excess interconnections, even though there were allowed many more fitness evaluations than the method using aggressive pruning.

In this contribution, we used DE to evolve real-valued Hill cofficients in systems of deterministic differential equations that represented the dynamics of GRNs. However, the concepts of (a) first evolving the correct dynamics in a relatively dense network, (b) subsequently condensing the network by alternating aggressive pruning of excess interactions with continued evolutionary refinement, and (c) using previously identified networks as non-evolvable subnetworks in the subsequent evolution of more complex, modular GRNs are easily generalized to other types of network representations (such as stochastic differential equations or graph-based representations) and other types of evolutionary algorithms.

Our interest is primarily in designing GRN circuits for synthetic biology. We envision using the approach described herein for identifying small GRN circuits that achieve specific desired dynamical behavior, to inform synthetic biologists as they seek to construct actual GRNs from DNA. While we used deterministic simulations in this work, stochastic simulations could be used to find GRNs that are robust to noise or that exhibit noise-driven behaviors. We also believe our approach could prove useful in other domains. For example, using a fitness function that tries to match the behavior of evolving GRNs with real time-series gene expression patterns (e.g., as in (Cao et al. 2010, Ruskin and Crane 2010)) our approach could prove useful in inferring naturally-occurring GRNs. Additionally, our approach could be used for evolving parsimonious topologies for Artificial Neural Networks.

In summary, we present a top-down, evolutionary approach that can be recursively applied to evolve increasingly complex, modular GRNs with few, if any, excess interactions. The approach uses DE to find dense GRNs that exhibit desired dynamical behaviors, interlaced with a procedure that aggressively prunes away excess interactions, favoring parsimonious GRNs even at the expense of small temporary degradations in raw fitness. Using this method we were able to recover known minimal GRNs for a bistable switch and an oscillatory circuit. We subsequently evolved modular GRNs that incorporated these previously identified sub-circuits in various ways to generate more complex target behaviors.

**ACKNOWLEDGMENTS**
We thank J. Lindle for contributions to the initial development of the project and M. Dunlop for helpful comments on the manuscript.